\documentstyle[12pt]{article}

\newcommand{\sect}[1]{\setcounter{equation}{0}\section{#1}}

\def\bseq{\begin{subequation}}  % = 1a 1b
\def\eseq{\end{subequation}}
\def\bsea{\begin{subeqnarray}}  % = 1.1a 1.1b
\def\esea{\end{subeqnarray}}

\newcommand{\beq}{\begin{equation}}
\newcommand{\eeq}{\end{equation}}
\newcommand{\bea}{\begin{eqnarray}}
\newcommand{\eea}{\end{eqnarray}}
\newcommand {\non}{\nonumber}

\renewcommand{\a}{\alpha}

\renewcommand{\d}{\delta}
\newcommand{\th}{\theta}
\newcommand{\Th}{\Theta}

\newcommand{\di}{\partial}

\newcommand{\G}{\Gamma}

\newcommand{\k}{\kappa}

\renewcommand{\l}{\lambda}

\def\Mb{\kern 2pt\mathchoice
            {%displaystyle
             \vbox{\hrule width10pt height 0.4pt depth 0pt
                 \kern 1.2pt\hbox{\kern -2pt$\displaystyle M$}}}
            {%textstyle
                 \vbox{\hrule width10pt height 0.4pt depth 0pt
                 \kern 1.2pt\hbox{\kern -2pt$\textstyle M$}}}
            {%scriptstyle \kern 0.5pt
\vbox{\hrule width6pt height 0.4pt depth 0pt
                 \kern 1.0pt\hbox{\kern -2pt$\scriptstyle M$}}}
            {%scriptscriptstyle \kern 0.5pt
                 \vbox{\hrule width5pt height 0.4pt depth 0pt
                 \kern 0.8pt\hbox{\kern -2pt$\scriptscriptstyle M$}}}}

\def\Sb{\kern 2pt\mathchoice
            {%displaystyle
                 \vbox{\hrule width6pt height 0.4pt depth 0pt
                 \kern 1.2pt\hbox{\kern -2pt$\displaystyle S$}}}
            {%textstyle
                 \vbox{\hrule width6pt height 0.4pt depth 0pt
                 \kern 1.2pt\hbox{\kern -2pt$\textstyle S$}}}
            {%scriptstyle
                 \vbox{\hrule width3.5pt height 0.4pt depth 0pt
                 \kern 1.0pt\hbox{\kern -2pt$\scriptstyle S$}}}
            {%scriptscriptstyle
                 \vbox{\hrule width3pt height 0.4pt depth 0pt
                 \kern 0.8pt\hbox{\kern -2pt$\scriptscriptstyle S$}}}}

\def\Rb{\kern 2pt\mathchoice
            {%displaystyle
                 \vbox{\hrule width5.5pt height 0.4pt depth 0pt
                 \kern 1.2pt\hbox{\kern -2.5pt$\displaystyle R$}}}
            {%textstyle
                 \vbox{\hrule width5.5pt height 0.4pt depth 0pt
                 \kern 1.2pt\hbox{\kern -2.5pt$\textstyle R$}}}
            {%scriptstyle
                 \vbox{\hrule width3.5pt height 0.4pt depth 0pt
                 \kern 1.0pt\hbox{\kern -2.2pt$\scriptstyle R$}}}
            {%scriptscriptstyle
                 \vbox{\hrule width3pt height 0.4pt depth 0pt
                 \kern 0.8pt\hbox{\kern -2.2pt$\scriptscriptstyle R$}}}}

  \def\pp{{\mathchoice
          {%displaystyle
              \kern 1pt%
              \raise 1pt
              \vbox{\hrule width5pt height0.4pt depth0pt
                    \kern -2pt
                    \hbox{\kern 2.3pt
                          \vrule width0.4pt height6pt depth0pt
                          }
                    \kern -2pt
                    \hrule width5pt height0.4pt depth0pt}%
                    \kern 1pt
           }
            {%textstyle
              \kern 1pt%
              \raise 1pt
              \vbox{\hrule width4.3pt height0.4pt depth0pt
                    \kern -1.8pt
                    \hbox{\kern 1.95pt
                          \vrule width0.4pt height5.4pt depth0pt
                          }
                    \kern -1.8pt
                    \hrule width4.3pt height0.4pt depth0pt}%
                    \kern 1pt
            }
            {%scriptstyle
              \kern 0.5pt%
              \raise 1pt
              \vbox{\hrule width4.0pt height0.3pt depth0pt
                    \kern -1.9pt  %[e]=0.15pt
                    \hbox{\kern 1.85pt
                          \vrule width0.3pt height5.7pt depth0pt
                          }
                    \kern -1.9pt
                    \hrule width4.0pt height0.3pt depth0pt}%
                    \kern 0.5pt
            }
            {%scriptscriptstyle
              \kern 0.5pt%
              \raise 1pt
              \vbox{\hrule width3.6pt height0.3pt depth0pt
                    \kern -1.5pt
                    \hbox{\kern 1.65pt
                          \vrule width0.3pt height4.5pt depth0pt
                          }
                    \kern -1.5pt
                    \hrule width3.6pt height0.3pt depth0pt}%
                    \kern 0.5pt%}
            }
        }}

  \def\mm{{\mathchoice
                       {%displaystyle
                             \kern 1pt
               \raise 1pt    \vbox{\hrule width5pt height0.4pt depth0pt
                                  \kern 2pt
                                  \hrule width5pt height0.4pt depth0pt}
                             \kern 1pt}
                       {%textstyle
                            \kern 1pt
               \raise 1pt \vbox{\hrule width4.3pt height0.4pt depth0pt
                                  \kern 1.8pt
                                  \hrule width4.3pt height0.4pt depth0pt}
                             \kern 1pt}
                       {%scriptstyle
                            \kern 0.5pt
               \raise 1pt
                            \vbox{\hrule width4.0pt height0.3pt depth0pt
                                  \kern 1.9pt
                                  \hrule width4.0pt height0.3pt depth0pt}
                            \kern 1pt}
                       {%scriptscriptstyle
                           \kern 0.5pt
             \raise 1pt  \vbox{\hrule width3.6pt height0.3pt depth0pt
                                  \kern 1.5pt
                                  \hrule width3.6pt height0.3pt depth0pt}
                           \kern 0.5pt}
                       }}

\def\pd{{\kern0.5pt
                   + \kern-5.05pt \raise5.8pt\hbox{$\textstyle.$}\kern
0.5pt}}

\def\pmd{{\kern0.5pt
                  \pm \kern-5.05pt \raise6.3pt\hbox{$\textstyle.$}\kern1.5pt}}

\def\md{{\mathchoice
   {%displaystyle
      {{\kern 1pt - \kern-6.2pt \raise5pt\hbox{$\textstyle.$}\kern 1pt}}}
    {%textstyle
      {{\kern 1pt - \kern-6.2pt \raise5pt\hbox{$\textstyle.$}\kern 1pt}}}
    {%scriptstyle
      {\kern0.5pt - \kern-5.05pt \raise3.4pt\hbox{$\textstyle.$}\kern0.5pt}}
    {%scriptscriptstyle
      {\kern0.5pt - \kern-5.05pt \raise3.4pt\hbox{$\textstyle.$}\kern0.5pt}}}}

\newcommand{\grad}{\nabla}

\def\Sc{\scriptstyle}

\newcommand{\reff}[1]{(\ref{#1})}

\newcommand{\shalf}{{\Sc\frac{1}{2}}}

\newcommand{\half}{\frac{1}{2}}

\renewcommand{\thefootnote}{\fnsymbol{footnote}}

\begin{document}

\newpage
\begin{titlepage}
\begin{flushright}
{hep-th/9801203}\\
{WATPHYS-TH-98/01}\\
{McGill/98-01}\\
\end{flushright}
\vspace{2cm}
\begin{center}
{\bf {\large}Super Black Hole from Cosmological Supergravity 
with a Massive Superparticle}\\
\vspace{1.5cm}
Marcia
 E. Knutt-Wehlau\footnote{knutt@physics.mcgill.ca}\\
\vspace{1mm}
{\em Physics Department, McGill University, Montreal, PQ
CANADA H3A 2T8}\\
\vspace{4mm}
and\\
\vspace{4mm}
R. B. Mann\footnote{mann@avatar.uwaterloo.ca} \\
\vspace{1mm}
{\em Physics Department, University of Waterloo, Waterloo, ON CANADA  N2L 3G1}

\vspace{1.1cm}
{{ABSTRACT}}
\end{center}

\begin{quote}

We describe in superspace a classical theory 
of two dimensional $(1,1)$ cosmological dilaton supergravity 
coupled to a massive
superparticle.  We give  an exact non-trivial superspace solution
 for the compensator superfield that 
describes the supergravity, and 
 then use this solution to construct a model of a two-dimensional
supersymmetric black hole.
\end{quote}

\vfill

\begin{flushleft}
January 1998

\end{flushleft}
\end{titlepage}

\newpage

\renewcommand{\thefootnote}{\arabic{footnote}}
\setcounter{footnote}{0}
\newpage
\pagenumbering{arabic}

\sect{Introduction}

There are very few exact classical non-trivial
solutions to supersymmetric field theories.
 As is elaborated on in \cite{sparticle},  even more rare are exact 
superspace supergravity solutions.  For classical supergravity theories,
one looks for {\it non-trivial} solutions -- those that cannot
be reduced by infinitesimal supersymmetry transformations to purely bosonic
solutions -- using the method given in \cite{aichel}.  However, it is possible
 to sidestep this
issue by examining classical supergravity problems in superspace \cite{bible}.  A
 {\em bona fide} superspace supergravity solution -- one which satisfies the
constraints -- has non-trivial torsion,  a
supercovariant quantity, and as such its value ultimately 
remains unchanged under a suitable
gauge transformation. Hence an exact superspace supergravity 
solution must necessarily
be non-trivial in this sense.  Approaching classical supergravity
problems from the superspace viewpoint obviates the triviality question. It is 
this approach that we take in this paper.

Following our previous motivation \cite{sparticle},
 we consider $(1,1)$ dilaton supergravity
with a cosmological constant,
coupled to a massive superparticle  in $(1+1)$ dimensions. 
 The supergravity part of the theory is  a supersymmetric
generalization of the $(1+1)$ dimensional ``R=T'' theory \cite{r3}.
This theory has the unique feature that the dilaton superfield
decouples
from the classical equations of motion, so that supermatter induces
superspace curvature, and the superspace curvature reacts back on 
the supermatter self-consistently. 
We obtain an exact  solution for the supergravity
compensator superfield  that completely 
describes the supergravity in superconformal
gauge, and use this 
compensator to
 construct a model of a supersymmetric black hole.

The outline of our paper is as follows. 
In section 2 we review bosonic cosmological dilaton gravity coupled to
a massive particle. In section 3, we outline cosmological dilaton $(1,1)$ 
supergravity coupled to a massive superparticle.
In section 4, we solve for the supergravity compensator
 in the presence of this superparticle, and in section 5, we discuss 
 the construction of a super black hole model using this
compensator.

\sect{Cosmological Dilaton Gravity}

 Before describing the supergravity action we use, we
briefly review the bosonic Lagrangian of a massive point particle
interacting with cosmological dilaton gravity \cite{tarasov},
 as it is simpler than the supersymmetric case, and 
illustrates the basic ideas.  

The action for $R=T$ theory is
\beq\label{act1}
S=S_G+S_M=\frac{1}{2\kappa}\int d^2x\left[{\sqrt{-g}}({\psi}R+{1\over 2}
(\nabla\psi)^2 )
-\kappa{\cal L}_M\right]
\eeq
where the gravitational coupling $\kappa = 8\pi G$. The action (\ref{act1})
ensures that the dilaton field $\psi$ decouples from the classical
equations of motion which, after some manipulation, are
\begin{eqnarray}
R&=&\kappa T_\mu^\mu \label{rt}\\
\frac{1}{2}\left(\grad_\mu\psi\grad_\nu\psi -\frac{1}{2}(\grad\psi)^2\right)
-\grad_\mu\grad_\nu\psi+g_{\mu\nu}\grad^2\psi &=& \kappa T_{\mu\nu} 
\end{eqnarray}
where $T_{\mu\nu} =\frac{1}{\sqrt{-g}} \frac{\delta {\cal L}_M}
{\delta g^{\mu\nu}}$ is 
the stress-energy tensor
and $R_{\mu\nu} = \partial_\lambda\Gamma^\lambda_{\mu\nu}-
\partial_\nu\Gamma^\lambda_{\mu\lambda}-
\Gamma^\lambda_{\mu\sigma}\Gamma^\sigma_{\lambda\nu}
+\Gamma^\lambda_{\lambda\sigma}\Gamma^\sigma_{\mu\nu}$ is our convention
for the Ricci tensor.

We take the matter Lagrangian to be that of a point particle
in a spacetime with non-zero cosmological constant 
\beq\label{matLag}
{\cal L}_M = -\sqrt{-g}\Lambda + 2m\int d\tau 
  \sqrt{- g_{\alpha\beta}\frac{dz^\alpha}{d\tau}\frac{dz^\beta}{d\tau}}
       \delta^{(2)}(x-z(\tau))
\eeq
so that
\beq\label{ptstress}
T_{\mu\nu} = \frac{1}{2}g_{\mu\nu}\Lambda + m\int d\tau \frac{1}{\sqrt{-g}} 
   g_{\mu\alpha}g_{\nu\beta}\frac{dz^\alpha}{d\tau}\frac{dz^\beta}{d\tau}
       \delta^{(2)}(x-z(\tau))
\eeq
is the relevant stress energy, with $z^\mu(\tau)$ being the worldline of the 
particle.

Choosing a frame at
rest with respect to the particle, the trace of the stress energy is
\beq\label{btrace}
T_\mu^\mu = -2m e^{-\rho}\delta(x-x_0) + \Lambda
\eeq
in conformal coordinates where $ds^2=e^{2\rho}dx^+ dx^-=e^{2\rho}(-dt^2+dx^2)/4$, 
with the location of the particle at $x=x_0$.
The
field equations (\ref{rt}) then become
\beq\label{lioupart}
\rho''(x) = -\frac{\kappa}{8}\Lambda e^{2\rho} + M e^\rho \delta(x-x_0)
\eeq
with $M= 2\pi Gm$.

Setting $a^2=\frac{\kappa}{8}|\Lambda|$, equation (\ref{lioupart}) has 
for $\Lambda >0$ the solution
\beq\label{lpartsol1}
\rho = -\ln\left(\cosh(a|x-x_0|+b)\right)
\eeq
where $\sinh(b) = \frac{M}{2a}$. If $\Lambda < 0$,
 the solution for $M<2a$ is
\beq\label{lpartsol2a} 
\rho = -\ln\left(\cos(a|x-x_0|+b)\right)
\eeq 
where $\sin(b) = \frac{M}{2a}$, or 
\beq\label{lpartsol2b}
\rho = -\ln\left(\sinh(b-a|x-x_0|)\right)
\eeq
if $M>2a$, where $\cosh(b)=\frac{M}{2a}$.  

In Schwarzschild-type coordinates the metric in the $\Lambda>0$ case
(\ref{lpartsol1}) becomes
\beq\label{lpart1}
ds^2 = -(-\frac{\kappa}{2}\Lambda Y^2+2M|Y|+1)dT^2
+ \frac{dY^2}{-\frac{\kappa}{2}\Lambda Y^2+2M|Y|+1}
\eeq
whereas for $\Lambda<0$ the solutions
(\ref{lpartsol2a}) and (\ref{lpartsol2b}) can respectively be transformed into
\beq\label{lpart2a}
ds^2 = -(\frac{\kappa}{2}|\Lambda| Y^2 + 2M|Y|+1)dT^2
+ \frac{dY^2}{\frac{\kappa}{2}|\Lambda| Y^2 + 2M|Y|+1}
\eeq
and
\beq\label{lpart2b}
ds^2 = -(\frac{\kappa}{2}|\Lambda| Y^2 + 2M|Y|-1)dT^2
+ \frac{dY^2}{\frac{\kappa}{2}|\Lambda| Y^2 + 2M|Y|-1}
\eeq
The metric (\ref{lpart2b}) is that of an anti de Sitter black hole
with mass parameter $M$. This solution, along with (\ref{lpart1})
 and (\ref{lpart2a}),
has been discussed previously in ref. \cite{tarasov}.

\sect{Cosmological $(1,1)$ Dilaton Supergravity}

We now extend the previous theory to a superspace formulation of
$(1,1)$ dilaton supergravity in two dimensions
with a cosmological constant, $L$. We use light-cone coordinates
$(x^\pp, x^\mm)= \half (x^1 \pm x^0)$ and $(\th^+, \th^-)$. In superconformal
gauge, the action  is
given by
\beq\label{scos}
I_C = -\frac{2}{\k}\int d^2x d^2\th (D_+ \Phi D_- \Phi + 4 \Phi D_- D_+ S
       - 4 e^{-2S} L)
\label{ccosmo}
\eeq
where $\Phi$ is the dilaton superfield,
 $S$ is the scalar compensator superfield that completely describes the 
supergravity in superconformal gauge, and the flat
supersymmetry covariant derivatives are given by
$D_\pm = (\di_+ + i \th^+
\di_\pp,  \di_- + i \th^- \di_\mm)$. 
We  simply list the results here, but details can be found in
\cite{sparticle}.  
The equations of motion are
\bea
 \Phi  &=& - 2 S\label{eqphi} \\
 D_- D_+ S &=&  e^{-2S} L 
\label{mateq}
\eea
from which it is clear that the dilaton decouples from the
theory, and that we recover the superLiouville equation for the 
compensator, $S$.

To obtain the component form of the superspace action \reff{ccosmo}, we identify
the components of the superfields by theta expansion (dropping the 
fermionic fields), 
\bea
S &=& -\frac{1}{2}\rho + \sigma \th^+\th^- \label{Scomp}\\
\Phi &=& -\frac{1}{2}\psi + \varphi \th^+\th^- \label{Phicomp}
\eea
and eliminate the auxiliary fields $\varphi,\sigma$ via their
 equations of motion. This yields 
\beq
I_C = \frac{1}{2\kappa}\int d^2x \left[-4\psi \partial_\pp\partial_\mm\rho
+\partial_\mm\psi \partial_\pp\psi - 16 L^2 e^{2\rho}  \right]
\eeq
for the component action. 
This is equivalent to (\ref{act1}) (using (\ref{matLag}) with $m=0$) provided 
$\Lambda = -\frac{32}{\k}L^2$.  Since $L$ must be real, this implies
 that from the superspace action, only the component action for anti de Sitter 
spacetimes is recovered. 
Alternatively, inserting the superfield expansions (\ref{Scomp},\ref{Phicomp})
into eqs. (\ref{eqphi}) and (\ref{mateq}) in the static case
yields after some manipulation
\beq\label{sconfrt}
\rho''(x) = 4 L^2 e^{2\rho}
 = -\frac{\kappa}{8}\Lambda e^{2\rho}
\eeq
which is equation  (\ref{lioupart}) with $M=0$.

We consider now extending the action (\ref{scos}) to include a
 massive superparticle.
We use $z = (x, t, \th)$ as the coordinates of the superspace, and ${z_0}(t) =
( {x_0}(t), {\th_0}(t))$ as the coordinates of the superparticle.
The technical details leading up to the action (\ref{spartact})
 below are the same
as in \cite{sparticle}, so we supply only the results here.

 The action for the superparticle in superconformal gauge is
\bea\label{spartact}
I_P &=&  2m \int dt dx d^2 \th \left\{ g^{-1} e^{-4S} \left[\shalf
(1+\dot{{x_0}}) + i {\th_0}^+
            \dot{{\th_0}}^+ \right]
            \left[\shalf ( 1 -\dot{{x_0}}) + i {\th_0}^- \dot{{\th_0}}^-\right]
\right. \non \\
            &+& i \left[ \shalf ( 1 +\dot{{x_0}}) + i {\th_0}^+ \dot{{\th_0}}^+
\right] D_+G_+
                 + i \left[ \shalf ( 1 -\dot{{x_0}}) + i {\th_0}^-
\dot{{\th_0}}^- \right] D_-G_-  \non\\
&+& \left. \dot{{\th_0}}^+ G_+  + \dot{{\th_0}}^- G_- + \frac{g}{4}
\right\} \d (x-{x_0}(t))
            \d (\th^+ - {\th_0}^+(t)) \d (\th^- - {\th_0}^-(t)) \label{part2}
\eea
where $g$ is the einbein on the 
worldline of the superparticle, and $G_\a \equiv e^S \G_\a$.
   The general gauge 
superfield  $\G_\a$  necessarily appears in the Wess-Zumino type term in 
the massive superparticle action in order for  consistent coupling
of the flat superparticle to supergravity.
Requiring that the supergravity constraints be satisfied introduces
 a constraint on the gauge field $G$, and we include
this constraint in the supergravity action by means
 of a lagrange multiplier, $\l$. 
Consequently, the dilaton supergravity part of the action is affected and
becomes
\bea
I_C &=& -\frac{2}{\k}\int d^2x d^2\th [D_+ \Phi D_- \Phi + 4 \Phi D_- D_+ S 
   - 4 e^{-2S} L \non \\
  &+& \k \l e^{-2S} (D_+G_- + D_-G_+ - i e^{-2S})] 
 \label{condil}
\eea

  {}From the sum of \reff{part2} and \reff{condil},
 we obtain for the equation of motion for $S$
\bea
 D_- D_+ S(z)  &-&  e^{-2S} L  \non \\
   &=& \frac{\k m}{2} \int dt' \left\{ g^{-1} e^{-4S} \left[\shalf
(1+\dot{{x_0}})
         + i {\th_0}^+ \dot{{\th_0}}^+ \right]
            \left[ \shalf ( 1 -\dot{{x_0}}) + i {\th_0}^-
\dot{{\th_0}}^-\right] \right\} \d^4 (z - {z_0}(t'))
      \non\\
&=& \frac{\k m}{4} \sqrt{\pi^2} e^{-2S} \d (x-{x_0}(t)) \d (\th^+ - {\th_0}^+(t))
    \d(\th^- - {\th_0}^-(t))
\label{comp}
\eea
where $ \sqrt{\pi^2} = \shalf \sqrt{1-\dot{{x_0}}^2}$ 
for a free particle.
Once we obtain the solution for $S$, it is possible to solve the constraint 
on $G$, but we shall not present this here.

\sect{Solution for Compensator}

 To solve for the compensator $S$ that describes the supergravity generated
by a superparticle in the presence of a cosmological constant, we consider
the superparticle to be stationary and fixed at ($x_0, \th_0$).
   In this case,
\reff{comp} becomes
\beq   
e^{2S} D_- D_+ S(z)  - L  
= \frac{M}{2} \d (x - x_0) \d (\th^+ - \th_0^+ ) \d(\th^- - \th_0^-)
\label{sliou}
\eeq
 where $M = \frac{\k m}{4}$ as before,  
and $\sqrt{\pi^2} = \shalf$ for $\dot{{x_0}}= 0$.
 We rewrite the equation in terms of $T = e^{2S}$
\beq
T D_+ D_- T - D_+T D_-T = -2T(L + \frac{M}{2} \delta(x- x_0)
\delta^{(2)}(\th-\th_0))\label{eqnT}
\eeq

   We solve the equation now for $T$ by analogy with 
the previous bosonic solution,
and also by experience with the form of the compensator in the $L=0$ case
\cite{sparticle}.  We find that just as in the bosonic case, the solution 
can be chosen  either  as
\beq
T(x, \th) = 2L(\th^+ - \th_0^+)(\th^- - \th_0^-)
               +  \cos[2L|X|+ c]
   \label{trigT}
\eeq
with $c = - \sin^{-1}(\frac{M}{4L})$,
or as 
\beq
T(x, \th) = 2L(\th^+ - \th_0^+)(\th^- - \th_0^-)
               +  \sinh[c-2L|X|]
   \label{hypT}
\eeq
with $c = \cosh^{-1}(\frac{M}{4L})$. Note that in the former
case $M<4L$, whereas in the latter case $M>4L$.
In these expressions,
\bea
|X| &\equiv& |x-x_0 - i(\th^+ \th_0^+ + \th^- \th_0^-)| \\ 
 &=&  |x-{x_0}| - i(\th^+ {\th_0}^+ + \th^-
{\th_0}^-)[\Th
(x-{x_0}) -\Th ({x_0}-x)]
     + 2 \th^+ \th^- {\th_0}^+ {\th_0}^- \d (x-{x_0}) \non
\eea
is to be understood as
a Taylor series expansion, and $\Th(x-{x_0})$ is the Heaviside function. 
We note that these are {\it specific} non-trivial solutions. The
 most general non-trivial solution to this problem
will be presented elsewhere \cite{bigsc}.
   
  Although we obtained the solution for $S$ assuming the particle was 
held fixed at $(x_0, \th_0)$, we note that the derivatives of $S$ with 
respect to the particle coordinates all vanish when
 evaluated at the particle position.
 This is sufficient to show that the ``force" on the particle 
due to the supergravity fields is zero, and hence this solution is in 
fact a solution to the full coupled equations of motion.

\sect{Discussion}

The solution (\ref{hypT}) is the supersymmetric analogue of the solution
(\ref{lpart2b}), and can be regarded as an anti de Sitter super black hole
in two-dimensional (1,1) superspace.  This solution is written in
the superconformal coordinates $z=(x, \th)$; to facilitate comparison with the 
results of \cite{tarasov},  we transform now to superspace coordinates 
$w=(u, \l)$ that correspond to Schwarzschild gauge, in which the dyad
of the bosonic subspace takes the form
\bea\label{2dzweibos}
{e_m}^a = \left[
\begin{array}{cc}
\sqrt{\a} & 0  \\
0 & {\sqrt{\a}}^{-1}
\end{array}
\right]
\eea

We find that the transformation from superconformal to Schwarzschild 
coordinates for $T$ of \reff{hypT} is
\bea
(c-2L|X|) &=& -\coth^{-1}[ \frac{4L}{\sqrt{\frac{M^2}{16L^2}+1}}
(|U| + U_0)]  \label{hyptrans}
\eea
where
\bea\label{Udef}
|U| & \equiv & |u-u_0 - i(\l^+ \l_0^+ + \l^- \l_0^-)|  \\
   & = & |u-{u_0}| - i(\l^+ {\l_0}^+ + \l^-{\l_0}^-)[\Th
     (u-{u_0}) -\Th ({u_0}-u)]
     + 2 \l^+ \l^- {\l_0}^+ {\l_0}^- \d (u-{u_0}) \non
\eea
and where $U_0$ is a constant.  One can choose $u_0=0$ without
loss of generality.
It is straightforward to work out the explicit relationship
between $(x,\theta^\pm)$ and $(u,\l^\pm)$, and also to compute 
explicit expressions for the gravitini, but we shall not do that here.
The transformation (\ref{hyptrans}) is actually valid only outside the 
event horizon for sufficiently large $|u-u_0|$. 
However, once one has the expression in Schwarzschild
coordinates it is easy to continue across the event horizon in
a manner analogous to that for the solution (\ref{lpart2b}). The 
compensator associated with the solution (\ref{hypT}) transformed 
via (\ref{hyptrans}) can then be used to compute the full vielbein associated
with the super black hole.

We can perform a similar transformation to facilitate comparison
between (\ref{lpart2a}) and its supersymmetric analogue (\ref{trigT}).
The former corresponds to the spacetime of a bosonic particle in
anti de Sitter space, and the latter is its superspace counterpart.
Here the supercoordinate transformation that takes us from
 $z=(x, \th)$ to $w=(u, \l)$ is given by 
\bea
(2L|X|+c) &=& \tan^{-1}[ \frac{4L}{\sqrt{1-\frac{M^2}{16L^2}}}
(|U| + U_0)]  \label{trigtrans}
\eea

We have found  expressions for the supergravity compensator
 that completely determine
the vielbeins of a super anti de Sitter black hole
and of a point particle in super anti de Sitter space. 
As with the bosonic case, the super black hole solution can only be obtained
provided $M>4L$.   A further exploration
of these solutions will be given in ref. \cite{bigsc}.

\noindent {\bf Acknowledgments}

This research was supported in part by the Ontario-Qu\'{e}bec Projects 
of Exchange at the University Level, NSERC of Canada,
and an NSERC Postdoctoral Fellowship.


\begin{thebibliography}{99}


\bibitem{sparticle}M.E. Knutt-Wehlau and R.B. Mann, hep-th/9708126, to be 
published in Nucl. Phys. B; see also references therein.

\bibitem{aichel} P.C. Aichelburg, Phys. Lett. B{\bf 91} (1980) 382.

\bibitem{bible}S. J. Gates, Jr., M.T. Grisaru, M. Ro\v{c}ek, and W. Siegel,
{\it Superspace},
Benjamin/Cummings, Reading, MA 1983.

\bibitem{r3}R.B. Mann, { Found. Phys. Lett.} {\bf 4}
(1991) 425; R.B. Mann, { Gen. Rel. Grav.} {\bf 24} (1992) 433.

\bibitem{tarasov} R.B. Mann, A. Shiekh and L. Tarasov, Nucl. Phys. B {\bf 341}
(1990) 134.

\bibitem{bigsc} M.E. Knutt-Wehlau and R.B. Mann, work in progress.


\end{thebibliography}
\end{document}